\documentclass[prb,aps,twocolumn]{revtex4}
\usepackage{graphicx,color,latexsym}
\usepackage{dcolumn}
\usepackage{amsmath,amssymb,epsf,bm}
\begin{document}
\title{Fraunhofer oscillations of the critical current at a varying Zeeman field in a spin-orbit coupled Josephson junction}
\author{A.~G. Mal'shukov}
\affiliation{Institute of Spectroscopy, Russian Academy of Sciences, Troitsk, Moscow, 108840, Russia}
\affiliation{Moscow Institute of Physics and Technology, Institutsky per.9, Dolgoprudny, 141700 Russia}
\affiliation{National Research University Higher School of Economics, Myasnitskaya str. 20, Moscow, 101000 Russia}
\begin{abstract}
 The Zeeman interaction  results in spontaneous current through a Josephson contact with  a spin-orbit coupled normal metal, even in the absence of any voltage, or phase bias. In the case of the Rashba spin orbit coupling of electrons in a two-dimensional (2D) electron gas this effect takes place for the Zeeman field which is parallel to the 2D system and to superconducting contacts. At the same time, the spontaneous current is absent when this field is perpendicular to the contacts. It is shown that in the latter case it may manifest itself in oscillations of the critical Josephson current at the varying Zeeman energy. These oscillations have a form of the Fraunhofer diffraction pattern. The Josephson current under the phase bias was calculated based on the semiclassical Green functions for a disordered 2D electron gas with the strong spin orbit coupling, as well as for surface electrons of a three dimensional topological insulator. In the latter case the diffraction pattern was found to be most pronounced, while in the Rashba gas the oscillations of the critical current are weaker.
\end{abstract}
\maketitle

\section{Introduction}
The interplay of the Zeeman and spin orbit interactions gives rise to a number of unusual physical phenomena in superconducting systems. One of the most spectacular observed \cite{Assouline} effects is the spontaneous supercurrent through a superconductor-normal metal-superconductor (SNS) Josephson junction (JJ), even in the absence of a phase bias \cite{Krive,Reinoso,Zazunov,ISHE,Liu,Yokoyama,Konschelle}. This current is induced by the so called anomalous phase $\varphi_0$ which enters into the Josephson current $J$ additively with the phase bias $\varphi$, so that $J=J_c\sin(\varphi+\varphi_0)$. In turn, the anomalous phase originates from a combined effect of the spin orbit coupling (SOC) and the Zeeman interaction in the normal metal. Usually, due to interface effects  SOC is strong in two-dimensional (2D) electronic systems. In most cases this coupling is represented by the Rashba SOC. \cite{Rashba} In this situation $\varphi_0$ is finite when the Zeeman field is parallel to 2D gas. This field may be created either by an external magnetic field, or by the exchange interaction of conduction electrons with magnetically polarized spins of impurities, as well as an adjacent magnetic insulator. $\varphi_0$  depends on the orientation of the Zeeman/exchange field in the 2D plane. Namely, it is finite if this field is perpendicular to the Josephson current. In contrast, $\phi_0=0$ if the Zeeman/exchange field is parallel to $\mathbf{J}$. Such an anisotropy takes place in the case of a standard geometry of the SNS contact, where the current direction is uniform inside the normal metal, as, for example in Ref. [\onlinecite{Assouline}].

The anomalous phase in JJ is a manifestation of a more general phenomenon. As was shown by Edelstein, \cite{Edelstein}  the Zeeman interaction leads to the coordinate dependent phase $\theta(\mathbf{r})$ of the Cooper pair wave function in spin-orbit coupled superconductors. This phase gives rise to helix variations of the superconductor order parameter \cite{Edelstein,Samokhin,Barzykin,Agterberg,Kaur,Agterberg2,Dimitrova} and to spontaneous currents around magnetic islands. \cite{Malsh island,Pershoguba,Hals} In the case of the Rashba SOC $\theta(\mathbf{r})$ tends to vary in the direction, which is perpendicular to the Zeeman field. Therefore, if the latter is perpendicular to superconducting contacts in JJ, as in Fig.1, the phase vary in the direction which is parallel to them ($y$-direction) and, hence, it cannot induce the current $J_x$ through the junction. However, it leads to spatial oscillations of  the proximity induced pairing amplitude. Such oscillations are caused by reflections of electrons from lateral edges of 2D normal metal in JJ which has a finite width along the $y$-axis. It will be shown that such a quantum interference effect results in oscillations of the critical Josephson current at the varying Zeeman energy. These oscillations form a pattern similar to the Fraunhofer diffraction.  This sort of the critical current dependence is often observed as a function of the external magnetic field which induces Meissner currents and a spatially dependent condensate phase in superconducting leads (see e.g. Ref.[\onlinecite{Suominen}]). The physics of this effect is quite different from that considered in the present work, because the latter takes account of the Zeeman energy, while orbital effects of the magnetic field are ignored. More complicated oscillations of the critical current are expected when the Zeeman field is non-uniform in the $y$-direction. For example, the magnetization may change its sign, if a domain wall (DW) is present inside JJ. It will be shown that depending on the position of  DW between two lateral edges these oscillations can vary from nonperiodic to Fraunhofer-like, with the doubled oscillation period, when DW is just in the middle of the junction.

This problem will be considered for a disordered normal 2D metal, or a doped semiconductor, which are in a weak contact to massive superconducting leads, as it is shown in Fig.1. The Josephson current will be calculated by using the semiclassical theory of electron Green functions. We will consider two models of 2D electrons. One of them is an electron gas with the parabolic band and the strong Rashba SOC. Another model is a Dirac system which represents the surface state of a doped topological insulator. In the latter case a possible effect of Majorana states on the Josephson current \cite{Fu} is ignored, mostly because in the considered tunneling regime for a strongly disordered system their role, as well as very existence present a separate problem.

The article is organized in the following way. In Sec.II the Usadel equations and boundary conditions for the model under consideration are formulated. In Sec.III the expression for  the Josephson current is derived and Sec.IV contains numerical results and their discussion.

\begin{figure}[tp]
\includegraphics[width=8 cm]{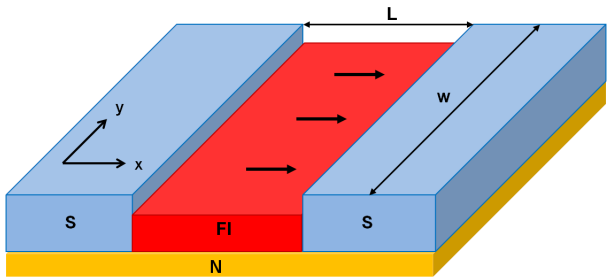}
\caption{(Color online) Sketch of the Josephson junction. Two superconductors (S), whose order parameter phases are $\varphi$ and $-\varphi$, are in a contact with a 2D normal metal (N) which is characterized by the strong Rashba spin-orbit coupling. The ferromagnetic insulator (FI) induces the Zeeman splitting $E_Z$ of electron energies in the normal metal (the direction of the Zeeman field is shown by arrows). In combination with the Rashba coupling this field leads to quantum interference of Cooper pair transmission amplitudes. As a  result, in a junction of the finite width $w$ the Josepson current oscillates as a function of $E_Z$. In the case of the uniform $E_Z$ these oscillations form the Fraunhofer diffraction pattern.}
\end{figure}
\section{Usadel equations and the boundary conditions}

The Josephson current through an SNS contact can be written in terms of the electron Green function of the normal metal which is calculated from the Schr\"{o}dinger equation with appropriate boundary conditions at the interface of the normal metal with superconducting contacts. The calculations are strongly simplified in the semiclassical approximation when all relevant length scales are much larger than the electron wavelength at the Fermi surface. An appropriate tool is represented by the Eilenberger equation  for  the semiclassical Green function. \cite{Larkin semiclassical,Eilenberger,Rammer,Serene} The latter is defined as
\begin{equation}\label{g}
\hat{g}_{\omega_n}(\mathbf{n}_{\mathbf{k}},\mathbf{r})=\frac{i}{\pi}\int d\xi_{\mathbf{k}} G_{\omega_n}(\mathbf{k,r})\,,
\end{equation}
where  $\mathbf{n}_{\mathbf{k}}=\mathbf{k_F}/k_F$.  $G_{\omega_n}(\mathbf{k,r})$ is obtained  from the Matsubara function $G_{\omega_n}(\mathbf{r,r}^{\prime})$  by Fourier transform with respect to $\mathbf{r}-\mathbf{r}^{\prime}$, and by setting $(\mathbf{r}+\mathbf{r}^{\prime})/2 \rightarrow \mathbf{r}$. Note, that $\hat{g}_{\omega_n}(\mathbf{n}_{\mathbf{k}},\mathbf{r})$  depends only on the direction of $\mathbf{k}$, which magnidude  is fixed on the Fermi line. $\hat{g}_{\omega_n}(\mathbf{n}_{\mathbf{k}},\mathbf{r})$ is a 2$\times$2 matrix in the spin space. When SOC is strong  it is convenient to use the helical basis which is formed by eigenstates of the matrix $\bm{\sigma}\times\mathbf{n}_{\mathbf{k}}$, where $\bm{\sigma}=(\sigma_x,\sigma_y,\sigma_z)$ is a vector of  Pauli matrices. The eigenvalues of this matrix are given by the helicity $\nu=\pm 1$. Accordingly, in the 2D electron system there are two bands with the energies $\xi_{\mathbf{k}}^{\nu}=k^2/2m + \nu \alpha k - \mu$, where $\alpha$ is the Rashba SOC constant and $\mu$ is the chemical potential. Each band crosses the Fermi energy ($\xi_{\mathbf{k}}^{\nu}=0$) at the corresponding wave-number $k_F^{\nu}$. The Fermi velocity $v_F=\sqrt{2\mu/m + \alpha^2}$ is the same in both bands, while densities of states $N^{\nu}_F=(m/2\pi)(1-\nu\alpha/v_F)$ are different. For a Dirac system one should set $m\rightarrow \infty$, so that only a single helical band crosses the Fermi surface, with $\nu=+$, or $-$, depending on the sign of $\mu$.

At the strong SOC a difference between $k_F^{+}$ and $k_F^{-}$ becomes larger than other relevant inverse length scales, such as  $E_Z/v_F$, $\Delta/v_F$ and $1/l$, where $l$, $E_Z$ and $\Delta$ are the mean free path of electrons, the Zeeman energy and the order parameter of  superconducting contacts, respectively. In this situation the approach can be used which was suggested for strongly Rashba coupled superconductors in Ref.[\onlinecite{Houzet}] and for Dirac systems in Refs.[\onlinecite{Zyuzin,Bobkova,Linder}]. This approach is based on the fact that at the large SOC the matrix elements of $\hat{g}_{\omega_n}(\mathbf{n}_{\mathbf{k}},\mathbf{r})$, which are nondiagonal in the helicity,  are small. Therefore, in the leading approximation only diagonal terms should be taken into account. Accordingly,  $\hat{g}_{\omega_n}(\mathbf{n}_{\mathbf{k}},\mathbf{r})$ in Eq.(\ref{g}) is represented by two functions $\hat{g}^{+}$ and $\hat{g}^{-}$. Each of them is associated with the integration over $\xi$ near respective Fermi circles of two helical bands. These functions can be written in the form
\begin{equation}\label{gnu}
\hat{g}^{\nu}_{\omega_n}(\mathbf{n}_{\mathbf{k}},\mathbf{r})=\frac{1}{2}g^{\nu}_{\omega_n}(\mathbf{n}_{\mathbf{k}},\mathbf{r})(1+\nu\bm{\sigma}\times\mathbf{n}_{\mathbf{k}})\,.
\end{equation}
So defined functions $g^{\pm}_{\omega_n}(\mathbf{n}_{\mathbf{k}},\mathbf{r})$ are $2\times2$ matrices in the Nambu space. They satisfy the normalization condition $g^{\nu 2}=1$.

Due to elastic scattering of electrons on impurities $g^{\nu }_{\omega_n}(\mathbf{n}_{\mathbf{k}},\mathbf{r})$ are almost isotropic in the $\mathbf{k}$-space. Therefore, in the leading  approximation they can be represented as
\begin{equation}\label{gbold}
g^{\nu}_{\omega_n}(\mathbf{n}_{\mathbf{k}},\mathbf{r})=g^{\nu}_{\omega_n}(\mathbf{r})+\mathbf{n}_{\mathbf{k}}\cdot\mathbf{g}^{\nu }_{\omega_n}(\mathbf{r})\,,
\end{equation}
where $\mathbf{g}\ll 1$, while $g^{\nu 2}_{\omega_n}(\mathbf{r})=1$. The isotropic functions $g^{\nu}_{\omega_n}(\mathbf{r})$ satisfy the Usadel \cite{Usadel} equation. The latter is
valid when the elastic scattering rate $1/\tau$ is much larger than $\Delta$ and $E_Z$, while  $l$ is much less than the length $L$ and the width $w$ of the junction. In Ref.[\onlinecite{Houzet}] a closed Usadel equation has been written for the spin-independent function
\begin{equation}\label{g0}
g_{0\omega_n}(\mathbf{r})=\frac{1}{2}(g^{+}_{\omega_n}(\mathbf{r})+g^{-}_{\omega_n}(\mathbf{r}))\,.
\end{equation}
This equation has the form
\begin{equation}\label{Usadel}
D\tilde{\bm{\nabla}}(g_0\tilde{\bm{\nabla}}g_0)=[\omega_n\tau_3+\gamma D F^2\tau_3g_0\tau_3,g_0] \,,
\end{equation}
where $\tilde{\bm{\nabla}}*=\bm{\nabla}*+i\mathbf{F}[\tau_3,*]$ and  $\mathbf{F}=(E_Z/v)(\mathbf{n}_Z\times\mathbf{e}_z)$, with $v=(\alpha^2+v_F^2)/2\alpha$. The unit vector $\mathbf{n}_Z$ denotes the direction of the Zeeman field which is parallel to the $xy$ plane, while $\mathbf{e}_z$ is the unit vector parallel to the $z$-axis. The diffusion coefficient $D=\tau(\alpha^2+v_F^2)/2$ and the dimensionless parameter $\gamma=(v_F^2-\alpha^2)/(2\alpha^2)$. Eq.(\ref{Usadel}) takes place also in TI \cite{Zyuzin,Bobkova,Linder}, where $\gamma=0$ and $D=\tau v_F^2$. Due to the vanishing $\gamma$ in TI it is possible in some cases to avoid the destructive effect of the magnetic field on the  superconducting proximity effect.

Let us assume that the Zeeman field is uniform inside the normal metal. In this case, by applying the unitary transformation $ g_0= U^{-1} \tilde{g}_0 U$, where $U=\exp(i\tau_3 \mathbf{F}\cdot\mathbf{r})$,  Eq.(\ref{Usadel})  can be transformed to
\begin{equation}\label{Usadel2}
D\bm{\nabla}(\tilde{g}_0\bm{\nabla}\tilde{g}_0)=[\omega_n\tau_3+\gamma D F^2\tau_3\tilde{g}_0\tau_3,\tilde{g}_0] \,,
\end{equation}
In the case of $\gamma=0$, as in TI,  the Zeeman field is removed from this equation, but not from the problem, because one should take into account boundary conditions (BC). Eq.(\ref{Usadel}) should be appended by BC on the boundaries with the left and right superconductors at $x=-L/2$ and $x=L/2$, respectively, as well as on the lateral edges of the junction at $y=0$ and $y=w$. In the case of diffusive electron transport the boundary conditions  at $x=\pm L/2$ can be written in the form, which is a straightforward  generalization of BC obtained by Kupryanov and Lukichev:\cite{Kupriyanov}
\begin{equation}\label{BC1}
Dg_0\tilde{\nabla}_xg_0|_{x=\pm L/2}=\pm\Gamma [g_0,g_s]|_{x=\pm L/2}\,,
\end{equation}
where $\Gamma$ is the tunneling parameter, which can be expressed \cite{Kupriyanov} through transparencies of contacts.  The same tunneling  parameters are assumed for both interfaces. $g_s|_{x=\pm L/2}\equiv g_s^{\pm}$ are the Green functions of the left and right superconducting contacts. They are assumed massive enough to ignore perturbations from the 2D normal metal. Therefore, these functions are fixed in the form of unperturbed semiclassical Green functions
\begin{equation}\label{gs}
g_s^{\pm}=e^{\pm i\tau_3 \varphi/2 }\frac{\tau_3\omega_n+\tau_2 \Delta}{\sqrt{\omega_n^2+\Delta^2}}e^{\mp i\tau_3 \varphi/2 }\,,
\end{equation}
where $-\varphi$ and $\varphi$ are the order parameter phases in the left and right contacts, respectively. In terms of the unitary transformed Green function Eq.(\ref{BC1}) takes the form
\begin{equation}\label{BC2}
D\tilde{g}_0\nabla_x\tilde{g}_0|_{x=\pm L/2}=\pm\Gamma [\tilde{g}_0,Ug_sU^{-1}]|_{x=\pm L/2}\,.
\end{equation}

At the lateral boundaries $\Gamma = 0$ and the boundary conditions for $\tilde{g}_0$ have the simple form
\begin{equation}\label{BC3}
D\tilde{g}_0\nabla_y\tilde{g}_0|_{y=0,w}=0\,.
\end{equation}

\section{Josephson current}
\subsection{Homogeneous magnetization}

The analytic solution of Eq.(\ref{Usadel2}) with boundary conditions Eq.(\ref{BC2}) and  Eq.(\ref{BC3}) can be obtained at the weak proximity effect, by assuming the small  tunneling parameter $\Gamma$. In this case one may represent the Green function in the form $\tilde{g}_0=\tau_3(\omega_n/|\omega_n|)+\delta\tilde{g}_0$ and linearize Eq.(\ref{Usadel2}) with respect to small $\delta\tilde{g}_0$. Further, linearized  Eq.(\ref{BC3}) can be resolved by the Fourier transform
\begin{equation}\label{Fourier}
\delta\tilde{g}_0(x,y)=\frac{1}{2}\sum_m \delta\tilde{g}_{0m}(x)\cos(q_m y)
\end{equation}
where $q_m=\pi m/w$ and $\delta\tilde{g}_{0m}$ is even with respect to $m\rightarrow -m$ . From linearized Eq.(\ref{Usadel2})  the functions $\delta\tilde{g}_{0m}(x)$ can be written in the form
\begin{equation}\label{gm}
\delta\tilde{g}_{0m}(x)=A_me^{\kappa_m x}+B_me^{-\kappa_m x}\,,
\end{equation}
where
\begin{equation}\label{kappa}
\kappa_m=\sqrt{\frac{2|\omega_n|}{D}+q_m^2+4\gamma F^2}\,.
\end{equation}
Within the linear approximation one can set in the right-hand side of  Eq.(\ref{BC2})  $\tilde{g}_0=\tau_3\omega_n/|\omega_n|$. As a result, the coefficients $A_m$ and $B_m$ in Eq.(\ref{gm}) can be easy found in the form
\begin{eqnarray}\label{AB}
A_m&=&\frac{\Gamma}{D\kappa_m}\left(\frac{\tilde{g}_{sm}^{+}-\tilde{g}_{sm}^{-}}{\cosh\frac{\kappa_m L}{2}}+\frac{\tilde{g}_{sm}^{+}+\tilde{g}_{sm}^{-}}{\sinh\frac{\kappa_m L}{2}}\right)\,,\nonumber\\
B_m&=&-\frac{\Gamma}{D\kappa_m}\left(\frac{\tilde{g}_{sm}^{+}-\tilde{g}_{sm}^{-}}{\cosh\frac{\kappa_m L}{2}}-\frac{\tilde{g}_{sm}^{+}+\tilde{g}_{sm}^{-}}{\sinh\frac{\kappa_m L}{2}}\right)\,,
\end{eqnarray}
where the Fourier coefficients $\tilde{g}_{sm}^{\pm}$ are given by
\begin{equation}\label{gsm}
\tilde{g}_{sm}^{\pm}=\frac{2}{w}\int_0^w dy e^{i\tau_3 \mathbf{F}\cdot\mathbf{r}^{\pm}}g_{s}^{\pm}e^{-i\tau_3 \mathbf{F}\cdot\mathbf{r}^{\pm}}\cos q_my\,,
\end{equation}
where $r^{\pm}_x=\pm L/2, r^{\pm}_y=y$. From Eqs.(\ref{gs}) and (\ref{gsm}) one can write the anomalous (nondiagonal) Green functions as
\begin{eqnarray}\label{gsm2}
\tilde{g}_{sm 12}^{\pm}&=&-F_y\frac{\Delta e^{\pm i(\varphi+F_xL) }}{w\sqrt{\omega_n^2+\Delta^2}}\frac{\left[e^{-2iF_yw}(-1)^m-1\right]}{F_y^2-q_m^2/4}\,,\nonumber \\
\tilde{g}_{sm 21}^{\pm}&=&-F_y\frac{\Delta e^{\mp i(\varphi+F_xL) }}{w\sqrt{\omega_n^2+\Delta^2}}\frac{\left[e^{2iF_yw}(-1)^m-1\right]}{F_y^2-q_m^2/4}\,.
\end{eqnarray}
The Fourier coefficients $\delta\tilde{g}_{0m}(x)$ are calculated by substituting Eq.(\ref{gsm2}) into Eq.(\ref{AB}), and further in Eq.(\ref{gm}). These coefficients are needed for the calculation of the Josephson current.

In terms of the Green function $G_{\omega_n}(\mathbf{k,r})$ the Josephson current $J$ can be written in the form \cite{Kopnin}
\begin{equation}\label{J}
J=e\frac{i\pi k_BT}{2}\int_0^w dy\sum_{n,\mathbf{k}}\mathrm{Tr}[\hat{v}^x_{\mathbf{k}}G_{\omega_n}(\mathbf{k,r})]\,,
\end{equation}
where the velocity operator $\hat{v}^x_{\mathbf{k}}=(k_x/m) -\alpha\sigma_y$. Since $J$ does not depend on $x$, this coordinate may be chosen in Eq.(\ref{J}) arbitrary. By using Eq.(\ref{g}) $G_{\omega_n}(\mathbf{k,r})$ can be expressed in terms of the semiclassical Green function. At the same time, two helical Fermi surfaces (circles) must be taken into account. At these circles $v^x_{\mathbf{k}_F}=v_Fn^x_{\mathbf{k}}$ is the same for both helical bands. Accordingly, by taking into account Eq.(\ref{gnu}) and by calculating the trace over spin variables, Eq.(\ref{J}) is written as
\begin{eqnarray}\label{J2}
&&J=ev_F\frac{i\pi k_BT}{2}\int_0^w dy\int \frac{d\mathbf{n}_{\mathbf{k}}}{4\pi}\sum_{n}n^x_{\mathbf{k}}\times \nonumber\\
&&\mathrm{Tr}\left[\tau_3\left(N^{+}_F g^+_{\omega_n}(\mathbf{n}_{\mathbf{k}},\mathbf{r})+
N^{-}_F g^-_{\omega_n}(\mathbf{n}_{\mathbf{k}},\mathbf{r})\right)\right]\,.
\end{eqnarray}
It is seen from Eq.(\ref{J2}) that only asymmetric in $\mathbf{k}$ parts of the functions $g^{\pm}$ contribute to this equation. They are given by the second term in Eq.(\ref{gbold}). At the same time, from Ref. \cite{Houzet} it is possible to express $\mathbf{g}^{\pm}_{\omega_n}(\mathbf{r})$  through the spin independent function $g_0(\mathbf{r})$ which is given by  Eq.(\ref{g0}). By this way we obtain
\begin{eqnarray}\label{gbold2}
&&\int \frac{d\mathbf{n}_{\mathbf{k}}}{4\pi}n^x_{\mathbf{k}}\left(N^{+}_F g^+_{\omega_n}(\mathbf{n}_{\mathbf{k}},\mathbf{r})+
N^{-}_F g^-_{\omega_n}(\mathbf{n}_{\mathbf{k}},\mathbf{r})\right)=\nonumber\\
&&-\frac{2DN_F}{v_F}g_0\tilde{\nabla}_xg_0\,,
\end{eqnarray}
where $2N_F=N_F^+ + N_F^-$. With the help of  Eq.(\ref{BC1}), or Eq.(\ref{BC2}), the spectral current in the right-hand side of Eq.(\ref{gbold2}) may be expressed  through the tunneling parameter at the interface with a contact.  At the right interface ($x=L/2$) Eq.(\ref{J2}) is thus transformed to
\begin{equation}\label{J3}
J=ie\Gamma\pi N_F k_BT\int_0^w dy\sum_{n}\mathrm{Tr}\left[\tau_3[\delta\tilde{g}_0,Ug_s^+U^{-1}]\right]\,.
\end{equation}
This equation can be further written in terms of Fourier transformed Green functions, where Fourier components of $\delta\tilde{g}_0$ and $Ug_s^+U^{-1}$ are given by Eqs.(\ref{Fourier}-\ref{AB}) and  Eqs.(\ref{gsm}-\ref{gsm2}), respectively. After substitution them in Eq.(\ref{J3}) the Josephson current takes the form
\begin{equation}\label{Jfin}
J=J_c\sin(2\varphi+\varphi_0)\,,
\end{equation}
where $\varphi_0=2F_xL=4L\alpha E_Zn_Z^y/(v_F^2+\alpha^2)$ is the anomalous phase, which in diffusive superconductors with the weak SOC has been calculated in Refs.[\onlinecite{ISHE,Konschelle}]. It is proportional to the projection $n_Z^y$ of the Zeeman field onto the $y$-axis. At the same time,  the critical  current depends on $n_Z^x$ and is given by
\begin{eqnarray}\label{Jc}
&&J_c=e\frac{\pi\Gamma^2}{Dw} N_F k_BT\sum_{n,m}\left(\frac{\sin^2 F_yw}{(F_y^2-q^2_{m})^2\kappa_m\sinh(\kappa_mL)}+\right.\nonumber\\
&&\left.\frac{\cos^2 F_yw}{(F_y^2-q^2_{m+\frac{1}{2}})^2\kappa_{m+\frac{1}{2}}\sinh(\kappa_{m+\frac{1}{2}}L)}\right)\frac{\Delta^2F_y^2}{\Delta^2+\omega_n^2}
\end{eqnarray}

\begin{figure}[tp]
\includegraphics[width=10.5 cm]{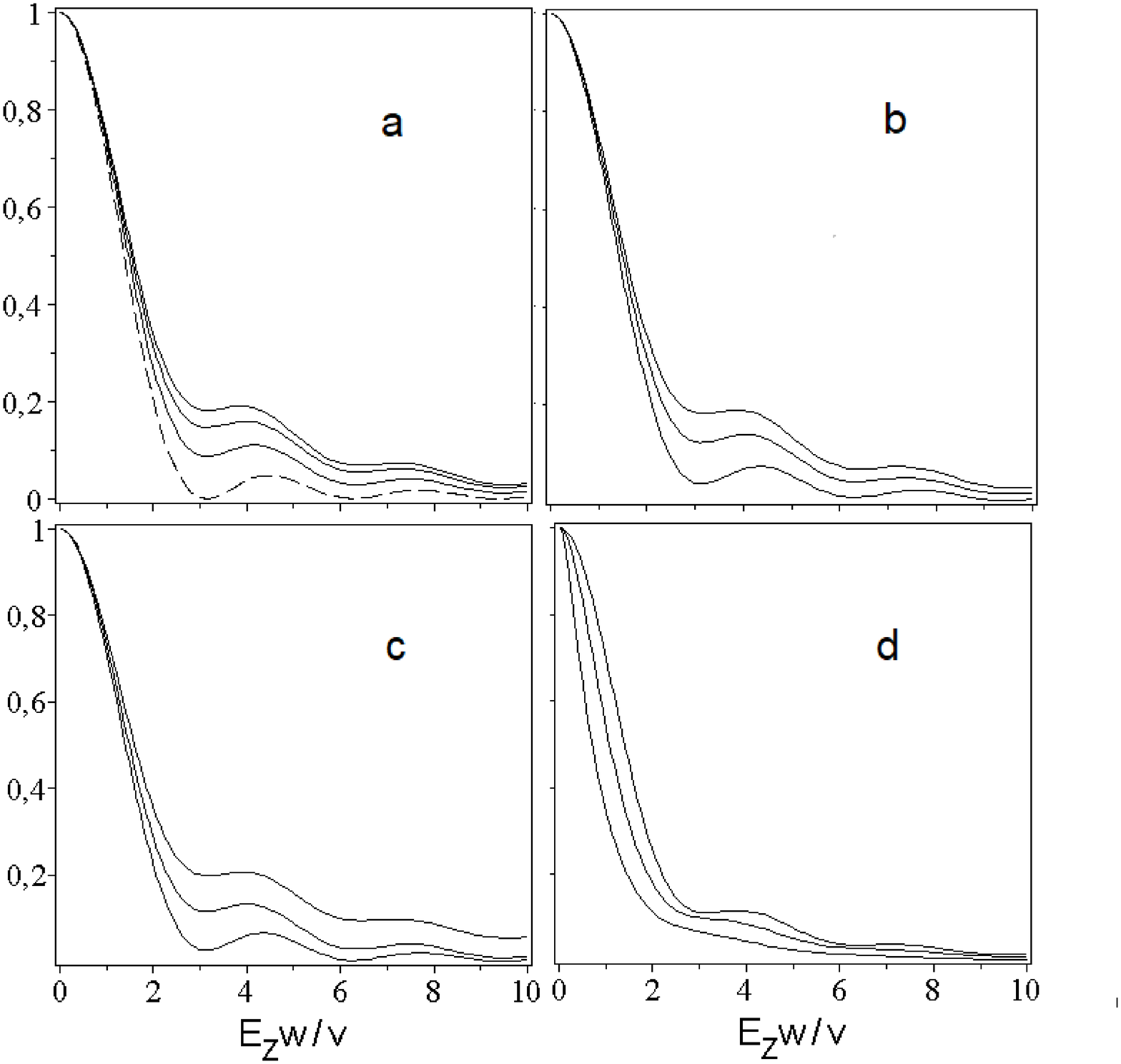}
\caption{Normalized Josephson current as a function of the Zeeman energy. a)  $\gamma=0$, $w/\xi=10$ and $L/w=0.1$. Curves from top to bottom, $\pi k_B T/\Delta=0.001,0.01$, and 0.03. For comparison, the dashed curve shows the Fraunhofer plot $\sin^2x/x^2$ where $x=E_Zw/v$. b)  $\gamma=0$, $L/w=0.1$ and $\pi k_B T/\Delta=0.03$. Curves from top to bottom: $w^2/\xi^2=10,50$, and 100. c) $\gamma=0$, $w/\xi=10$ and $\pi k_B T/\Delta=0.02$. Curves from top to bottom, $L/w=0.05, 0.2$, and 0,5. d) $w^2/\xi^2=50$, $L/w=0.1$, and $\pi k_B T/\Delta=0.02$.  Curves from top to bottom, $\gamma=0.1, 1$, and 5.}
\end{figure}

\subsection{Inhomogeneous magnetization. Domain wall}

As follows from Eq.(\ref{Jc}), the critical Josephson current does not depend on the sign of $F_y$ (the projection of the Zeeman field  on the $x$-axis). In this connection it is interesting to find out what happens when this projection  is not homogeneous. For example, it can change the sign, if there are  domain walls inside the JJ, such that $n_Z^x$ change its sign at $y=y_i$, where $y_i$ denote positions of domain walls, whose size is assumed to be much less than $w$.  It is evident that as compared to the homogeneous case one should not expect any change in the current, if quantum interference effects on the tunneling of Cooper pairs are ignored, because individually each domain results in  the same current. On the other hand, as seen from Eq.(\ref{Jc}), in the case of the single-domain magnetization the quantum interference results in oscillations of the current at varying  $F_y$. In a multidomain case one may expect a more complicated oscillation pattern due to the interdomain interference of transmission amplitudes. As a simple example, let us consider a single DW which is located just in the middle of the junction at $y=w/2$. In this situation one must take into account that $F_y$ in Eq.(\ref{gsm}) changes sign at $y=w/2$. Then, Eq.(\ref{gsm}) gives the same equation for $2g_{s(2m)}$ as Eq.(\ref{gsm2}), with $F_y$ substituted for $F_y/2$. The Josephson current is represented by the same equation as Eq.(\ref{Jc}) with $\kappa_m \rightarrow \sqrt{2(|\omega_n|/D)+4q_m^2+4\gamma F^2}$, instead of $\kappa_m$ given by Eq.(\ref{kappa}). Therefore, an evident result of DW is the doubling of the current oscillation period, as a function of $F_y$. This situation resembles the Fraunhoffer diffraction when the slit width is reduced by a factor of two. When the DW is located in an arbitrary point $y\neq w/2$ the current is not a periodic function of $F_y$ anymore. Therefore, the motion of DW through JJ, or reorientation of the domain magnetization are accompanied by transformations of the interference patterns between non-periodic and periodic ones, with different periodicities.

\section{results and discussion}

The results, which are presented at Figs.1(a-d) demonstrate the dependence of the Josephson critical current on the Zeeman energy. This current is normalized to $J_c$ at $E_Z=0$. With the higher Zeeman field the current decreases and oscillates. The oscillations are caused by quantum interference effects and resemble the Fraunhofer oscillations. For reference, the dashed line in Fig2a shows the Fraunhofer diffraction pattern.  In order to evaluate a relevant range of Zeeman energies, let us take $w=1000$ nm and $v=10^6$ m/s. Than, the parameter $wE_Z/v=\pi$ at $E_Z\simeq 2$meV. As long as the $g$-factor is not very high, this energy corresponds to rather strong external magnetic fields (tens of Tesla). At the same time, the exchange coupling with an adjacent magnetic insulator, or magnetically ordered impurities can provide such a strong Zeeman energy. The oscillations, as can be seen from the first three plots, are most pronounced when  $\gamma=0$. This parameter in Eq.(\ref{Usadel}) determines the suppression of the superconducting proximity effect by the magnetic field in the considered case of the strong SOC and the relatively weak Zeeman field. 

It should be noted that at such conditions the proximity effect is very different from the effect which usually takes place in ferromagnetic Josephson junctions with relatively weak SOC. In the latter case the crucial role is played by the long-range proximity effect caused by triplet Cooper pairs. The latter may be created either due to an inhomogeneity in the Zeeman field \cite{Bergeret,Kadigrobov}, or by a combined action of SOC and the  Zeeman field \cite{ISHE,Konschelle}. In the presence of SOC the triplet proximity effect extends to distances that are restricted by the Dyakonov-Perel \cite{DP} spin relaxation length \cite{ISHE}. These distances can be rather large, if the  spin-orbit coupling is not very strong. In contrast, in the considered here case this coupling is strong. It is much stronger than the Zeeman field and the elastic scattering rate, so that Cooper correlations take place at two Fermi circles corresponding to two helicities. Therefore, Cooper pair spins are locked to relative electron momenta, as seen in the pairing function Eq.(\ref{gnu}). As a result, the elastic impurity scattering leads to the relaxation of these spins together with momenta within the electron's mean free path, which is short in the considered diffusive approximation. An interplay of strongly mixed triplet and singlet Cooper pairs partly results in the gauge field in the covariant derivative of Eq.(\ref{Usadel}), while the second term of this equation gives rise to the depairing effect. It is interesting that this term vanishes ($\gamma=0$) in the case of a Dirac system, as it follows from Refs.[\onlinecite{Zyuzin,Bobkova,Linder}]. At the same time, for a Rashba coupled metal $\gamma$  is finite and increases when the ratio $\alpha/v_F$ decreases. Indeed, in Fig.2d the amplitude of Fraunhofer oscillations decreases with larger $\gamma$. Other parameters which control the current are  $L/w$ , $\pi k_BT/\Delta$, and the ratio $w/\xi$, where $\xi=\sqrt{D/2\Delta}$ is the coherence length. It is seen from a comparison between curves in Fig.2  that the oscillation amplitude increases with larger  $L/w$,  $\pi k_BT/\Delta$, and $w/\xi$.

The considered here interference effect of the Zeeman interaction on the critical Josephson current can be useful for increasing functionalities of devices which are based on hybrid magnetic-superconducting systems. In particular, this effect might play an important role in interaction of magnons with collective excitations of superconducting quantum circuits which integrate JJ of the  considered in this work type.

\emph{Acknowledgements} - The work was partly supported by the Russian Academy of Sciences program "Actual
 problems of low-temperature physics."


\end{document}